\documentstyle[preprint,eqsecnum,aps]{revtex}  % DON'T CHANGE
\begin{document}                % INITIALIZE - DON'T CHANGE%
\preprint{KNU-TH-25, \\ February \\1995}
\draft
\title{New Asymptotic Expansion Method for the Wheeler-DeWitt Equation
{\footnote
{\it To be published in Physical Review D {\bf 52} (1995)}}}
\author{Sang Pyo Kim\cite{Kim}}
\address{Department of Physics \\ Kunsan National University
\\ Kunsan 573-701, Korea}
% \author{}   % Use this and the next line only if there is a second
% \address{Another University, etc.}  % address. (Remove the left % marks)
%
\maketitle
\begin{abstract}                % DON'T CHANGE THIS LINE
A new asymptotic expansion method is developed
to separate the Wheeler-DeWitt equation into
the time-dependent Schr\"{o}dinger equation
for a matter field and the
Einstein-Hamilton-Jacobi equation for
the gravitational field including
the quantum back-reaction of
the matter field. In particular, the nonadiabatic basis
of the generalized invariant
for the matter field Hamiltonian
separates the Wheeler-DeWitt equation
completely in the asymptotic limit
of $m_p^2$ approaching infinity.
The higher order quantum corrections
of the gravity to the matter field
are found. The new asymptotic expansion method is valid
throughout all regions of superspace
compared with other expansion methods
with a certain limited region of validity.
We apply the new asymptotic
expansion method to the minimal
FRW universe.
\end{abstract}

\section{INTRODUCTION}                 % Introduction goes below

Recently the quantum field theory
in a curved spacetime for matter fields
has been studied in the context of quantum cosmology.
In quantum cosmology one derives
from the Wheeler-DeWitt (WDW) equation,
a tentative quantum gravity theory,
the quantum field theory for the matter
fields in the curved space which
is the time-dependent Schr\"{o}dinger equation,
a Tomonaga-Schwinger functional equation.
The time-dependent Schr\"{o}dinger equation for  matter
fields has an  advantage over the canonical quantization on the
solution space of wave equation for the matter fields
in that one can take into account the
higher order quantum corrections of
the gravity to the matter fields
and the quantum back-reaction
of the matter fields to the gravity.
Banks \cite{Banks} showed that
the semiclassical limit (WKB approximation)
of the WDW equation for the pure gravity
was the Einstein-Hamilton-Jacobi (EHJ)
equation for the gravity, which turned out
equivalent to the classical Einstein
equation \cite{Gerlach}. In Ref. \cite{Brout1}
the induced gauge potential
due to some cosmological mode was
considered in semiclassical gravity.
The semiclassical gravity was extended to
quantum cosmological models for the gravity coupled to
matter fields by including
the expectation value, a back-reaction,
of the matter fields to the EHJ equation
\cite{Kiefer1,Singh,Paz,Gundlach,Kiefer2}.
In a quantum cosmological model for the gravity
coupled to matter fields whose mass scale
is much smaller than the Planck mass,
there are not only the back-reaction of the matter fields to
the EHJ equation but also
the geometric phases to the quantum
states of the matter fields
\cite{Balbinot,Brout2,Datta,Kim1,Kiefer3}.

In this paper we develop a new
asymptotic expansion method which
separates the WDW equation for the
gravity coupled to a matter field
into the time-dependent Schr\"{o}dinger equation
for the matter field and the EHJ equation for
the gravity including the quantum back-reaction of
the matter field. When one expands with respect to some basis of
quantum states for the matter field and introduces a cosmological time
the WDW equation is equivalent
to a matrix equation which involves
the cosmological time-dependent elements.
In particular it is shown in
the nonadiabatic basis of the
eigenstates of the generalized
invariant for the matter field Hamiltonian
that the WDW equation is equivalent to
the matrix equation which consists
of dominant diagonal elements
and nondiagonal perturbation elements
proportional to the asymptotic parameter
$1/m_p^2$.
In other words we are able to separate the WDW equation
into the time-dependent matrix equation
for the matter field and the EHJ
equation with the back-reaction of the matter field.
We obtain the higher order quantum corrections of the
gravity to the matter field as a power series of
$1/m_p^2$. We also compare the
new asymptotic expansion method
with other related works. Finally,
we apply the new asymptotic expansion method to the minimal
FRW universe.

The organization of this paper
is as follows. In Sec. II we introduce
the new asymptotic expansion method for the WDW equation.
In Sec. III the higher order quantum
corrections of the gravity
to the matter field
are obtained. In Sec. IV we compare
the new asymptotic expansion method with
other methods used to derive the
time-dependent Schr\"{o}dinger equation
for the matter field. Finally in Sec. V
we apply the new asymptotic expansion
method to the Friedmann-Robertson-Walker
universe minimally coupled to
a free massive scalar field. The quantum
back-reaction of the matter field
and the EHJ equation are found explicitly
using the generalized invariant for a well-known
time-dependent harmonic oscillator.

\section{NEW ASYMPTOTIC EXPANSION METHOD}

With the fundamental constants
such as $c$ the speed of light, $\hbar$ the
Planck constant, and $G$ the gravitational constant,
inserted explicitly, the WDW
equation takes the form
\begin{equation}
\left[ -\frac{\hbar^2}{2m_p^2} G_{ab}
\frac{\delta^2}{\delta h_a \delta h_b}
 - 2 m_p^2c^2 \sqrt{h} ~{}^{(3)}R(h_a )
+ \hat{H}_m (\pi_\phi, \phi, h_a ) \right] \Psi (h_a, \phi) = 0,
\label{WD eq.}
\end{equation}
where $m_p^2 = c^2/32 \pi G$ is the Planck mass squared
and $G_{ab}$ is the symmetrized supermetric on
the superspace of the three-geometry.
There are two asymptotic parameters
$1/\hbar \rightarrow \infty $
and $m_p^2 \rightarrow \infty$ for the
WDW equation, which can be compared with just one asymptotic
parameter $1/ \hbar \rightarrow \infty$ for the Schr\"{o}dinger
equation in quantum mechanics.
The asymptotic limit $m_p^2 \rightarrow \infty$ can be considered
as the limiting case either of $G \rightarrow 0$
or $c^2 \rightarrow \infty$ or
both of them. We shall consider
here only the weak gravitational coupling
limit $G \rightarrow 0$ without any loss of generality.

The method in this paper is neither to evolve the wave functions
of the WDW equation from some Cauchy initial data
\cite{Kim2} nor to separate the matrix effective gravitational
field Hamiltonian equation by expanding the wave functions
with respect to some basis for the matter field Hamiltonian
\cite{Kim1}. The key point of our method is to separate
asymptotically the Schr\"{o}dinger equation for the
matter field by introducing a cosmological
time via the gravitational action which in turn
includes the quantum back-reaction of the matter field.
This can be achieved by setting the wave function in the form
\begin{equation}
\Psi (h_a, \phi) = \exp\left(\frac{i}{\hbar} S(h_a)\right)
\Phi (\phi,h_a).
\label{wave func.}
\end{equation}
Once we use the asymptotic parameter $1/\hbar \rightarrow \infty$
in the WDW equation just as for the WKB expansion in quantum
mechanics, then we obtain the following intermediate equation
\begin{eqnarray}
\Biggl[\frac{1}{2m_p^2} G_{ab} \biggl( 2i \hbar
\frac{\delta S}{\delta h_{(a}}
\frac{\delta}{\delta h_{b)}}   &-&  \frac{\delta S}{\delta h_a}
\frac{\delta S}{\delta h_b}
+ \hbar^2 \frac{\delta^2}{\delta h_a \delta h_b}
- i\hbar\frac{\delta^2 S}{\delta h_a \delta h_b}
\biggr) \nonumber\\
 &+&  2 m_p^2 c^2 \sqrt{h} ~{}^{(3)}R(h_a) -
\hat{H}_m (\pi_\phi, \phi, h_a)
\Biggr] \Phi(\phi, h_a ) = 0,
\label{intermediate eq.}
\end{eqnarray}
where $S$ is the gravitational action
to be determined consistently later on and $(a,b)$ denotes
symmetrization with respect to the indices $a$ and $b$.
Assuming that the prefactor
of the wave function in Eq. (\ref{wave func.})
oscillates and is peaked around a classical trajectory,
we may first introduce the cosmological
time along the classical trajectory in this
oscillatory region of the superspace by
\begin{equation}
\frac{\delta}{\delta \tau} \equiv \frac{1}{m_p^2}
G_{ab} \frac{\delta S}{\delta h_{(a}}
\frac{\delta}{\delta h_{b)}},
\label{cosmological time}
\end{equation}
and then rewrite Eq. (\ref{intermediate eq.}) as
\begin{eqnarray}
\Biggl[ i \hbar \frac{\delta}{\delta \tau} - \hat{H}_m (\pi_\phi, \phi, h_a)
&+& 2m_p^2 c^2 \sqrt{h} ~{}^{(3)}R(h_a)
 -  \frac{1}{2m_p^2} G_{ab} \frac{\delta S}{\delta h_a}
\frac{\delta S}{\delta h_b} \nonumber\\
&-&i\frac{\hbar}{2m_p^2}G_{ab}
\frac{\delta^2 S}{\delta h_a \delta h_b}
+  \frac{\hbar^2}{2m_p^2} G_{ab}
\frac{\delta^2}{\delta h_a \delta h_b}
 \Biggr] \Phi(\phi, h_a) = 0.
\label{Schr eq.}
\end{eqnarray}
It should be remarked that Eq. (\ref{Schr eq.}) can be interpreted as
the exact time-dependent Schr\"{o}dinger equation for the matter
field including the gravitational quantum back-reaction, since
if we neglect the last two terms in Eq. (\ref{Schr eq.}), which is valid
in the limit $1/m_p^2 \rightarrow 0$, it is nothing but
the conventional time-dependent Schr\"{o}dinger equation
\cite{Banks,Kiefer1,Halliwell}:
\begin{equation}
i\hbar \frac{\delta}{\delta \tau} \Phi (\phi, h_a)
= \hat{H}_m (\pi_\phi, \phi, h_a) \Phi (\phi, h_a)
\label{approx. Schr eq.}
\end{equation}
provided that the gravitational action satisfies
the conventional EHJ equation:
\begin{equation}
\frac{1}{2m_p^2} G_{ab} \frac{\delta S^{(0)}}{\delta h_a}
\frac{\delta S^{(0)}}{\delta h_b} - 2m_p^2 c^2 \sqrt{h} ~{}^{(3)}R(h_a) = 0
\label{approx. EHJ eq.}
\end{equation}
The second to the last term, $-i(\hbar/2m_p^2)G_{ab}
(\delta^2 S/\delta h_a \delta h_b)$, not only is asymptotically
small as a power of $1/m_p^2$ but also
violates the unitarity of matter field equation,
so we shall not consider this term in this paper.

Our method to be developed below differs
from others in that instead of taking the conventional
EHJ equation in Eq. (\ref{approx. EHJ eq.})
to obtain the conventional time-dependent Schr\"{o}dinger equation
in Eq. (\ref{approx. Schr eq.}) we treat the exact time-dependent
Schr\"{o}dinger equation in Eq. (\ref{Schr eq.}), expand the exact
quantum states of the matter field by some basis, and finally solve
the matrix equation equivalent to Eq. (\ref{Schr eq.}).
For the sake of simplicity, we shall use the ket-bra vector notation
for the quantum states of the matter field, and act
the bra vector on the ket vector to denote the inner product.
The exact quantum state in Eq. (\ref{Schr eq.}) for the matter field
can be expanded by some basis
$\left|\Phi_k (\phi,h_a) \right>$ which constitutes a
complete set of orthonormal vectors as
\begin{equation}
\Phi(\phi, h_a) = \sum_{k}^{\atop} c_k (h_a) \left| \Phi_k
(\phi, h_a )\right>,
\label{quantum state}
\end{equation}
where $c_k (h_a)$ are coefficient functions of the gravitational field only.
Both $c_k (h_a)$ and $\left|\Phi_k
(\phi,h_a) \right>$ are still unknowns
which will be determined systematically
later on. Then Eq. (\ref{Schr eq.}) is equivalent to the following
matrix equation
\begin{eqnarray}
i \hbar \frac{\delta}{\delta \tau} c_k (h_a) &+& \Omega^{(0)}(h_a) c_k (h_a)
\nonumber\\
&+& \sum_{n} {\atop} \left(- \Omega_{kn}^{(1)} (h_a) + \Omega_{kn}^{(2)}
(h_a) +
\frac{\hbar^2}{2m_p^2} \Omega_{kn}^{(3)} (h_a) \right) c_n (h_a) = 0,
\label{matrix eq.}
\end{eqnarray}
where
\begin{eqnarray}
\Omega^{(0)} (h_a)
&=& - \frac{1}{2m_p^2} G_{ab}
\frac{\delta S}{\delta h_a}
\frac{\delta S}{ \delta h_b}
+ 2 m_p^2 c^2 \sqrt{h} ~{(3)}R(h_a),
\nonumber \\
\Omega^{(1)}_{kn}(h_a)
&=& \left< \Phi_k ( \phi, h_a) \right|
\hat{H}_m ( \pi_\phi, \phi, h_a )
\left| \Phi_n ( \phi, h_a) \right>,
\nonumber \\
\Omega^{(2)}_{kn}(h_a)
&=& i \hbar \left< \Phi_k ( \phi, h_a) \right|
\frac{\delta}{\delta \tau} \left|
\Phi_n ( \phi, h_a) \right>,
\nonumber \\
\Omega^{(3)}_{kn} (h_a)
&=& G_{ab} \left< \Phi_k (\phi,h_a) \right|
\frac{\delta^2}{\delta h_a \delta h_b}
\left| \Phi_n (\phi, h_a) \right>
\nonumber\\
&+&G_{ab} \frac{\delta^2}{\delta h_a \delta h_b}
\delta_{kn}
- 2i ({\rm A}_{(a})_{kn}
\frac{\delta}{\delta h_{b)}},
\end{eqnarray}
where
\begin{equation}
(A_a)_{kn} = i
\left<\Phi_k(\phi,h_a) \right|
\frac{\delta}{\delta h_a}
\left|\Phi_n(\phi,h_a) \right>.
\end{equation}

The task of solving the WDW equation (\ref{WD eq.}) is now reduced to that
of solving the matrix equation (\ref{matrix eq.}). However,
in the asymptotic limit of $1/m_p^2 \rightarrow 0$, we may neglect the
last term of summation in Eq. (\ref{matrix eq.})
and get the approximate equation
\begin{equation}
i \hbar \frac{\delta}{\delta \tau} c_k (h_a)
+ \Omega^{(0)} c_k (h_a)
+ \sum_{n} {\atop}
\left(- \Omega_{kn}^{(1)} (h_a)
+ \Omega_{kn}^{(2)} (h_a) \right)
c_n (h_a) = 0.
\label{approx. matrix eq.}
\end{equation}
On the other hand, if one introduces a generalized
invariant obeying the following invariant equation
\begin{equation}
\frac{d}{d \tau} \hat{I} (h_a) = \frac{\partial}{\partial \tau}
\hat{I}(h_a) - \frac{i}{\hbar}
\left[\hat{I} (h_a) , \hat{H}_m (h_a) \right] = 0
\label{invariant eq.}
\end{equation}
for the matter field Hamiltonian, then there is a well-known decoupling
theorem \cite{Lewis} for the generalized invariant such that
\begin{equation}
\Omega_{kn}^{(1)} (h_a) = \Omega_{kn}^{(2)} (h_a)
\label{decoupling}
\end{equation}
for different eigenstates, $ k\neq n$, of the generalized invariant:
\begin{equation}
\hat{I}(h_a) \left| \Phi_k (\phi, h_a) \right> = \lambda_k \left| \Phi_k
(\phi,h_a) \right>.
\label{eigenstate}
\end{equation}
Since all of the off-diagonal equations in Eq. (\ref{approx. matrix eq.})
vanish, we have only the diagonal equations left
\begin{equation}
i\hbar \frac{\delta}{\delta \tau} c_k (h_a) + \left(
\Omega^{(0)}(h_a)  -
\Omega_{kk}^{(1)} (h_a) + \Omega_{kk}^{(2)} (h_a)\right) c_k (h_a) = 0.
\label{diagonal eq.}
\end{equation}
Recollecting that the gravitational action $S$ is still undetermined,
we may take a simple ansatz of the form
\begin{equation}
\Omega^{(0)} (h_a) - \Omega^{(1)}_{kk} (h_a) = 0,
\label{lowest order sol.}
\end{equation}
and
\begin{equation}
i \hbar \frac{\delta}{\delta \tau} c_k (h_a) +
\Omega_{kk}^{(2)} (h_a) c_k (h_a) = 0.
\label{zeroth order sol.}
\end{equation}
Equation (\ref{lowest order sol.}) is nothing but the familiar
EHJ equation with the quantum back-reaction of
the matter field included
\begin{equation}
\frac{1}{2m_p^2} G_{ab} \frac{\delta S}{\delta h_a} \frac{\delta S}{ \delta
h_b}  -  2m_p^2 c^2 \sqrt{h} ~{}^{(3)}R(h_a)
+ \left< \Phi_k ( \phi, h_a) \right| \hat{H}_m (\pi_\phi, \phi, h_a) \left|
\Phi_k (\phi, h_a) \right> = 0.
\label{EHJ eq.}
\end{equation}
It should be noted that the EHJ equation (\ref{EHJ eq.})
is an implicitly coupled
nonlinear equation in which the $k$th
eigenstate is defined by the generalized
invariant (\ref{invariant eq.}), which in turn is defined using
the cosmological time (\ref{cosmological time}) through the gravitational
action $S$. This means that the gravitational action does depend
on the mode-number $k$, i.e. $S_{(k)}$, and so does the cosmological time
$\tau_{(k)}$. The physical implication is that there are infinite number
of gravitational action $S_{(k)}$, where $k$ runs over all quantum numbers
of the generalized invariant which satisfies
Eq. (\ref{EHJ eq.}). The cosmological time $\tau_{(k)}$
is defined along each gravitational action $S_{(k)}$. It was shown
in Ref. \cite{Kim2} that there is a spectrum of infinite
number of wave functions that depend on the modes. The gravitational
action $S_{(k)}$ corresponds to the wave function $\Psi_{(k)}$
and the cosmological time is defined along the classical trajectory
around the peak of wave function. Now, the wave function
(\ref{wave func.}) becomes
\begin{equation}
\Psi_{(k)} (h_a,\phi) = \exp \left(\frac{i}{\hbar}S_{(k)}(h_a)\right)
\Phi_{(k)}(\phi,h_a)
\end{equation}
and so does the quantum state of the matter field
\begin{equation}
\Phi_{(k)}(\phi,h_a) = \sum_n c_n(h_a) \left|
\Phi_{(k)n}(\phi,h_a) \right>.
\end{equation}
We find the solution to Eq. (\ref{zeroth order sol.})
\begin{equation}
c_k^{(0)}(h_a) = d_k \exp \left(\frac{i}{\hbar} \int \Omega_{kk}^{(2)} (h_a)
d\tau_{(k)} \right),
\end{equation}
where $d_k$ is a constant.
It was implicitly assumed that
all the Eqs. (\ref{matrix eq.}) through (\ref{EHJ eq.})
should depend on the mode-number $(k)$; we shall, however,
work with the $(k)$ mode only and
drop the mode-number hereafter throughout this paper.

\section{HIGHER ORDER QUANTUM CORRECTIONS}

We now turn to the time-dependent Schr\"{o}dinger equation
(\ref{Schr eq.}) for the matter field
with the gravitational action
$S$ and thereby the cosmological time $\tau$ determined by
Eq. (\ref{EHJ eq.}) and Eq. (\ref{cosmological time}), respectively.
The time-dependent Schr\"{o}dinger equation, the gravitational
action and the cosmological time are all defined along the $(k)$th
mode, whose mode-number will be omitted.
Its matrix equivalent in Eq. (\ref{matrix eq.}) now satisfies
\begin{equation}
i \hbar \frac{\delta}{\delta \tau} c_k (h_a)
+ \Omega_{kk}^{(2)} (h_a) c_k (h_a)
+ \frac{\hbar^2}{2m_p^2}
\sum_{n}^{\atop} \Omega^{(3)}_{kn} (h_a) c_n (h_a) = 0.
\label{higher order eq.}
\end{equation}
By substituting
\begin{equation}
c_k(h_a) =  \exp \left(\frac{i}{\hbar} \int \Omega_{kk}^{(2)} (h_a)
d\tau \right) \tilde{c}_k(h_a)
\end{equation}
into Eq. (\ref{higher order eq.}) we obtain the following equation
\begin{equation}
i \hbar \frac{\delta}{\delta \tau} \tilde{c}_k (h_a)
+ \frac{\hbar^2}{2m_p^2}
\sum_{n}^{\atop} \tilde{\Omega}^{(3)}_{kn} (h_a) \tilde{c}_n (h_a) = 0,
\label{mod. higher order eq.}
\end{equation}
where
\begin{equation}
\tilde{\Omega}^{(3)}_{kn} (h_a) =
\exp \left(-\frac{i}{\hbar} \int \Omega_{kk}^{(2)} (h_a)
d\tau \right) \Omega_{kn}^{(3)} (h_a)
\exp \left(\frac{i}{\hbar} \int \Omega_{nn}^{(2)} (h_a)
d\tau \right).
\end{equation}
The simple ansatz (\ref{zeroth order sol.}) was a solution to
the asymptotic approximate diagonal equation (\ref{diagonal eq.}),
therefore should be regarded as the zeroth
order solution,
\begin{equation}
\tilde{c}_k^{(0)}(h_a) = d_k,
\end{equation}
to Eq. (\ref{higher order eq.}) with the EHJ equation
(\ref{EHJ eq.}) still satisfied.
We may find the solution for Eq. (\ref{mod. higher order eq.}) perturbatively
in a power series of $\hbar /2m_p^2$:
\begin{equation}
\tilde{c}_k = \sum_{n}{\atop} \left( \frac{\hbar}{2m_p^2} \right)^n
\tilde{c}_k^{(n)}.
\label{power series}
\end{equation}
where they satisfy the recursion equation
\begin{equation}
i \frac{\delta}{\delta \tau} \tilde{c}^{(l)}_k (h_a) + \sum_{n}{\atop}
\tilde{\Omega}^{(3)}_{kn} (h_a) \tilde{c}^{(l-1)}_n (h_a) = 0.
\label{recursion eq.}
\end{equation}
The solution to the recursion equation is the ordered {\it l}th multiple
integral
\begin{equation}
\tilde{c}^{(l)}_k = i^l \sum_{n_i, \\i=1,\ldots ,l} {\atop} \int
\tilde{\Omega}^{(3)}_{kn_1} (h_a)
\int \tilde{\Omega}^{(3)}_{n_1 n_2} (h_a) \cdots \int
\tilde{\Omega}^{(3)}_{n_{l-1} n_l} (h_a) \tilde{c}^{(0)}_{n_l}.
\label{series sol.}
\end{equation}
For an instance, when an initial data
$\tilde{c}_n^{(0)} = d_k \delta_{kn}$, $d_k$
= constant, is imposed, the quantum state of the matter
field including the gravitational correction up to the first order is
\begin{equation}
\Phi(\phi,H_A)  \sim d_k
\exp \left(\frac{i}{\hbar} \int \Omega_{kk}^{(2)} (h_a)
d\tau \right)
\left(\left|\Phi_k(\phi,h_a)\right>
+ i\frac{\hbar}{2m_p^2} \sum_n \int
\tilde{\Omega}_{kn}^{(3)}(h_a)d\tau \left|\Phi_n(\phi,h_a)
\right>\right).
\label{first order quantum state}
\end{equation}
Thus the first order transition to the other states
from the initially prepared state
is suppressed by a factor of $1/m_p^2$ because there comes
a factor of $\hbar$ from the integration and similarly the {\it n}th
order transition is suppressed by a factor of $(1/m_p^2)^n$.
It is to be noted that the procedure employed
in this paper is quite similar
to the perturbation method for a quantum system with a small
perturbation term; here the parameter for the
smallness of perturbation term is $\hbar /2m_p^2$ and
Eq. (\ref{series sol.}) is the perturbative series for the given
choice of the basis of eigenstates of the generalized invariant,
which is also the solution for the
unperturbed equation (\ref{approx. Schr eq.}).

\section{COMPARISON WITH OTHER RELATED WORKS}

In this section, we shall compare the result
of this paper with other related works.

First, we shall answer in part the question whether
the conventional EHJ equation (\ref{approx. EHJ eq.}) or the EHJ equation
(\ref{EHJ eq.}) is right through the investigation of a quantum
cosmological model.
Suppose a quantum cosmological model with the gravitational
super-Hamiltonian
\begin{equation}
\hat{H}_g (h_a) = \frac{1}{2m_p^2} G_{ab}
\hat{\pi}^a \hat{\pi}^b - 2m_p^2c^2
\sqrt{h} ~^{(3)}R(h_a).
\label{gravitation Ham.}
\end{equation}
When coupled to the matter field, the gravitational super-Hamiltonian
(\ref{gravitation Ham.}) leads to the WDW equation (\ref{WD eq.}).
Quantizing the gravitational Hamiltonian Eq. (\ref{gravitation Ham.})
by substituting $\hat{\pi}^a = (\hbar/i)(\delta /\delta
h_a)$, taking the wave function of the form $\Psi (h_a) \sim
\exp(iS(h_a)/\hbar)$, and keeping dominant terms only, we just obtain
the conventional EHJ equation
\begin{equation}
\frac{1}{2m_p^2}G_{ab} \frac{\delta S}{\delta h_a}
\frac{\delta S}{\delta h_b} - 2m_p^2 c^2 \sqrt{h} ~^{(3)}R(h_a) = 0,
\label{conv. EHJ eq.}
\end{equation}
which is the same as Eq. (\ref{approx. EHJ eq.}).
Finally, putting the conventional EHJ equation
(\ref{conv. EHJ eq.}) and neglecting
the last two terms in Eq. (\ref{Schr eq.}),
we obtain the time-dependent Schr\"{o}dinger equation
\begin{equation}
i\hbar \frac{\delta}{\delta \tau} \Phi (\phi,h_a) = \hat{H}_m (\pi_{\phi},
\phi, h_a) \Phi (\phi,h_a).
\label{time-dep. Schr eq.}
\end{equation}
We found the exact quantum states in terms of the eigenstate of the
generalized invariant \cite{Kim1}:
\begin{equation}
\Phi (\phi, h_a) = \sum_{k}{\atop} d_k \exp \left(\frac{i}
{\hbar} \int \Omega_{kk}^{(2)}(h_a) d\tau \right) |\Phi_k(\phi,h_a)>.
\label{time-dep. quantum state}
\end{equation}
It is to be noted that the conventional EHJ equation
(\ref{conv. EHJ eq.}) is equal to the approximation
$\Omega^{(0)}(h_a) = 0$ and the gravitational field-dependent
coefficients in Eq. (\ref{time-dep. quantum state}) can also be determined
by directly integrating Eq. (\ref{diagonal eq.}) with  both
$\Omega^{(0)}(h_a) = 0$ and $\Omega^{(1)}(h_a) = 0$
substituted. However, the lowest order solution
(\ref{zeroth order sol.}) of the coefficient function
does not depend on the gravitational
field, i.e., is a constant and the higher order solution
(\ref{series sol.}) depends on the gravitational field.
The significant difference on the
coefficient functions comes from
the fact that in this paper we have used the EHJ equation
(\ref{EHJ eq.}) with the quantum back-reaction of the matter field rather
than the conventional EHJ equation (\ref{conv. EHJ eq.}).
Below we put forth a criterion on whether Eq. (\ref{EHJ eq.})
or Eq. (\ref{conv. EHJ eq.}) should be used for the
correct gravitational action.

In order to show the relation between the EHJ equation (\ref{EHJ eq.})
and the conventional EHJ equation (\ref{conv. EHJ eq.}),
we expand the action (\ref{EHJ eq.}) perturbatively in the inverse
power of the Planck mass:
\begin{equation}
S_k = \sum_{n}^{\atop} m_p^{2(1-n)} S^{(n)}_k,
\label{gravitation action series}
\end{equation}
whose lowest order action obeys
\begin{equation}
O(m_p^2) : \frac{1}{2} G_{ab}
\frac{\delta S_k^{(0)}}{\delta h_a} \frac{\delta
S_k^{(0)}}{\delta h_b} - 2 c^2 \sqrt{h}~^{(3)}R(h_a) = 0,
\label{lowest grav. action}
\end{equation}
and the first two higher order actions obey
\begin{eqnarray}
O(m_p^0) &:& G_{ab} \frac{\delta S^{(0)}_k}{\delta h_{(a}} \frac{\delta
S_k^{(1)}}{\delta h_{b)}} - \Omega_{kk}^{(1)} + \Omega_{kk}^{(2)}
= 0, \nonumber\\
O(m_p^{-2} ) &:& G_{ab} \left( \frac{\delta S_k^{(0)}}{\delta h_{(a}}
\frac{\delta S^{(2)}_k}{\delta h_{b)}} + \frac{1}{2} \frac{\delta
S^{(1)}_k}{\delta h_a} \frac{\delta S^{(1)}_k}{\delta h_b} \right) = 0.
\label{higher order grav. action}
\end{eqnarray}
It is worthy to note that the lowest order contribution
(\ref{lowest grav. action}) gives nothing but the conventional EHJ
equation (\ref{conv. EHJ eq.}). So the above question
whether Eq. (\ref{EHJ eq.}) or Eq. (\ref{conv. EHJ eq.})
should be used is closely related to
the question whether the asymptotic expansion of the gravitational
action (\ref{gravitation action series}) gives the correct
gravitational action.

However, contrary to the belief widely accepted that the conventional EHJ
equation (\ref{conv. EHJ eq.}) gives the correct gravitational action,
we give a counter-example showing that the asymptotic
expansion (\ref{gravitation action series}) leads
to a wrong gravitational action.
In the case of the Friedmann-Robertson-Walker universe
minimally coupled to a scalar field with a power-law
potential, the WDW equation takes the form
\begin{equation}
\left[ \frac{\hbar^2}{2m_p^2} \frac{1}{a}
\frac{\partial^2}{\partial a^2} - 2
m_p^2c^2 a - \hbar^2 \frac{1}{a^3}
\frac{\partial^2}{\partial \phi^2} + 2 a^3
U(\phi) \right] \Psi (a,\phi) = 0.
\label{FRW WD eq.}
\end{equation}
from which it follows that $G_{aa} = -1/a$ and $\hat{H}_m (\phi, a) =
-\hbar^2 \partial^2/a^3 \partial \phi^2 +2a^3 U(\phi)$.
The EHJ equation (\ref{conv. EHJ eq.}) becomes
\begin{equation}
\frac{1}{2m_p^2} \frac{1}{a} \left( \frac{\partial S}{\partial a} \right)^2 +
2m_p^2c^2 a = 0.
\label{FRW conv. EHJ eq.}
\end{equation}
Direct integration by quadrature yields the gravitational action
$S(a) = \pm im_p^2ca^2$ and the wave function
\begin{equation}
\Psi (a, \phi ) = \exp\left(\pm \frac{1}{\hbar} m_p^2 ca^2 \right)
\Phi (\phi,a).
\end{equation}
Likewise, the lowest order gravitational action
(\ref{lowest grav. action}) also has the same value
$S^{(0)} = \pm ica^2$. Therefore, the cosmological time
Eq. (\ref{cosmological time}) leads to
an imaginary one $\tau = \pm i\ln (a/2c)$.
Both the conventional EHJ equation (\ref{conv. EHJ eq.}) and the dominant
term of the asymptotic expansion of the gravitational action
(\ref{gravitation action series}) always lead to the wave functions
with an exponential behavior due to the curvature term,
the second term in Eq. (\ref{FRW WD eq.}).
However, both in the adiabatic basis method \cite{Kiefer4} which expands
the wave functions by the gravitational field-dependent
eigenfunctions of the matter field Hamiltonian and in
the superadiabatic expansion method \cite{Kim2} in which transitions
among different eigenstates are taken into account during
the evolution of the Universe, the resulted wave functions
show not only the exponential behavior for a
large three-geometry but also the oscillatory behavior
for an intermediate three-geometry depending on the
quantum number of the matter field Hamiltonian.
The oscillatory behavior of wave functions is inevitable
to the classical Lorentzian universe such as the present
Friedmann-Robertson-Walker universe. Therefore, in the case of
the minimally coupled Friedmann-Robertson-Walker universe
it can be inferred that the conventional EHJ
equation (\ref{conv. EHJ eq.}) has a certain limited region of
the large three-geometry for the validity, whereas the EHJ
equation (\ref{EHJ eq.}) holds not only for the large three-geometry
prevailing with the curvature term but also for the intermediate
three-geometry prevailing with the quantum back-reaction of the matter field.
The EHJ equation (\ref{EHJ eq.}) should be used
in order to give the correct gravitational action
valid for all the regions of superspace.

Second, we shall compare the new asymptotic expansion method with
the adiabatic expansion method.
Quite similarly as in the new asymptotic expansion method
in which one expands the quantum state of matter field
by the nonadiabatic basis of the eigenstates of the generalized invariant,
in the adiabatic expansion method one expands the quantum state of the matter
field by the adiabatic basis of the instantaneous eigenstates
of the matter field Hamiltonian itself and includes the quantum
back-reaction of the matter field with respect
to the adiabatic basis. By
defining the instantaneous eigenstates
\begin{equation}
\hat{H}_m (h_a) \left|\Phi_k(\phi,h_a) \right>_{ad} = H_k(h_a)
\left|\Phi_k(\phi,h_a) \right>_{ad},
\end{equation}
and expanding the quantum state
\begin{equation}
\Phi (\phi,h_a) = \sum_{k}^{\atop} c_{ad,k}(h_a)
\left|\Phi_k(\phi,h_a) \right>_{ad},
\end{equation}
one obtains the matrix equation
\begin{eqnarray}
i \hbar \frac{\delta}{\delta \tau} c_{ad,k} (h_a) &+& \Omega^{(0)}(h_a)
c_{ad,k} (h_a) \nonumber\\
&+& \sum_{n} {\atop} \left(- \Omega_{ad,kn}^{(1)} (h_a) + \Omega_{ad,kn}^{(2)}
(h_a) +
\frac{\hbar^2}{2m_p^2} \Omega_{ad,kn}^{(3)} (h_a) \right) c_{ad,n} (h_a) = 0,
\label{ad. matrix eq.}
\end{eqnarray}
where
\begin{eqnarray}
\Omega^{(0)} (h_a)
&=& - \frac{1}{2m_p^2} G_{ab} \frac{\delta S}{\delta h_a}
\frac{\delta S}{ \delta h_b}
+ 2 m_p^2 c^2 \sqrt{h} ~^{(3)}R(h_a),
\nonumber \\
\Omega^{(1)}_{ad,kn}(h_a)
&=& {}_{ad}\left< \Phi_k ( \phi, h_a) \right|
\hat{H}_m ( \pi_\phi, \phi, h_a ) \left|
\Phi_n ( \phi, h_a) \right>_{ad},
\nonumber \\
\Omega^{(2)}_{ad,kn}(h_a)
&=& i \hbar ~{}_{ad}\left< \Phi_k ( \phi, h_a)
\right|
\frac{\delta}{\delta \tau} \left|
\Phi_n ( \phi, h_a) \right>_{ad},
\nonumber \\
\Omega^{(3)}_{ad,kn} (h_a)
&=& G_{ab}~ {}_{ad}\left< \Phi_k (\phi,h_a)\right|
\frac{\delta^2}{\delta h_a \delta h_b}
\left| \Phi_n (\phi, h_a) \right>_{ad}
\nonumber\\
&+&G_{ab} \frac{\delta^2}{\delta h_a \delta h_b}
 \delta_{kn}
- 2i ({\rm A}_{(ad,a})_{kn}
\frac{\delta}{\delta h_{b)}},
\end{eqnarray}
where
\begin{equation}
(A_{ad,a})_{kn} = i
{}_{ad}\left<\Phi_k(\phi,h_a) \right|
\frac{\delta}{\delta h_a}
\left|\Phi_n(\phi,h_a) \right>_{ad}.
\end{equation}
Again, in the asymptotic limit of $1/m_p^2 \rightarrow 0$,
we may neglect the
last term in Eq. (\ref{ad. matrix eq.}) and get the approximate equation
\begin{equation}
i \hbar \frac{\delta}{\delta \tau} c_{ad,k} (h_a) + \Omega^{(0)}(h_a)
c_{ad,k} (h_a)
+ \sum_{n} {\atop} \left(- \Omega_{ad,kn}^{(1)} (h_a) + \Omega_{ad,kn}^{(2)}
(h_a) \right) c_n (h_a)  = 0.
\label{ad. approx. matrix eq.}
\end{equation}
It should, however, be remarked that in the adiabatic
expansion method the off-diagonal elements of
 the coupling matrix
do not vanish, $\Omega^{(1)}_{kn} \neq \Omega^{(2)}_{kn}$
for $k \neq n$, in strong
contrast with those in the nonadiabatic expansion method.
Therefore, one should
solve the whole adiabatic approximate matrix equation
(\ref{ad. approx. matrix eq.})
with the elements of the coupling matrix
accounting for  transition between different eigenstates
instead of the frequently used adiabatic diagonal equation in
Eq. (\ref{ad. approx. matrix eq.})
\begin{equation}
i\hbar \frac{\delta}{\delta \tau} c_k (h_a) + \left( \Omega^{(0)}(h_a) -
\Omega_{kk}^{(1)} (h_a) + \Omega_{kk}^{(2)} (h_a) \right) c_k (h_a) = 0.
\label{ad. diagonal eq.}
\end{equation}
Because $\Omega^{(1)}_{ad,kn}$ for $k \neq n$, have an order of magnitude
comparable to $\Omega^{(1)}_{ad,kk}$, it is not justified to use
the adiabatic diagonal equation (\ref{ad. diagonal eq.}),
which is frequently used in literatures
under the assumption that the off-diagonal elements be neglected.
{}From Eq. (\ref{ad. diagonal eq.}) one also obtains the frequently used
EHJ equation with the adiabatic quantum back-reaction of
the matter field
\begin{eqnarray}
\frac{1}{2m_p^2} G_{ab} \frac{\delta S}{\delta h_a} \frac{\delta S}{ \delta
h_b}  -  2m_p^2 c^2 \sqrt{h} ~{}^{(3)}R(h_a)
 + {}_{ad}\left< \Phi_k ( \phi, h_a) \right| \hat{H}_m (\pi_\phi, \phi, h_a)
\left|
\Phi_k (\phi, h_a) \right>_{ad} = 0.
\label{ad. EHJ eq.}
\end{eqnarray}
Here we have the same subtlety as in Sec. II  that the EHJ
equation with the adiabatic quantum back-reaction
depends on the mode-number $k$. So we have
the gravitational action $S_{(k)}$ and
the cosmological time $\tau_{(k)}$. The difference between the
EHJ equation (\ref{ad. EHJ eq.}) and the EHJ equation (\ref{EHJ eq.})
is that in the former
the quantum back-reaction of the matter field is explicitly
given, whereas in the latter it is implicitly determined
via the cosmological time which is defined by the gravitational
action.

Third, there has been a recent study of geometric phases
as a mechanism for the asymmetry of the cosmological time
\cite{Kim1}. In particular, in the basis of eigenstates
of the generalized invariant one may define
a gauge potential (Berry connection)
\begin{equation}
A_a (h_a) = i \vec{U}^* (\phi, h_a) \frac{\delta}{\delta h_a} \vec{U}^T
(\phi, h_a),
\label{gauge potential}
\end{equation}
where
\begin{equation}
\vec{U} (\phi,h_a) = \left( \begin{array}{c}
|\Phi_0 (\phi, h_a)> \\
|\Phi_1 (\phi, h_a)> \\
\vdots \\
|\Phi_n (\phi, h_a)> \\
\vdots
\end{array} \right),
\label{column vector}
\end{equation}
is a column vector, and the asterisk ${}^*$
and the superscript ${}^T$ denote dual
and transpose operations, respectively.
One can show that
\begin{equation}
\frac{\delta}{\delta \tau} \vec{U} (\phi, h_a) = -i \frac{1}{m_p^2} G_{ab}
\frac{\delta S}{\delta h_{(a}} A^T_{b)} \vec{U} (\phi, h_a),
\end{equation}
and
\begin{equation}
G_{ab} \frac{\delta}{\delta h_a} \frac{\delta}{\delta h_b} \vec{U} (\phi,
h_a) = G_{ab} \left( \frac{\delta A^T_b}{\delta h_a} + (A_a A_b)^T \right)
\vec{U} (\phi, h_a).
\label{second diff.}
\end{equation}
Then the matrix equivalent (\ref{matrix eq.}) to the WDW equation
is entirely determined
by the gravitational action and the gauge potential.
It should be remarked again that the matter field Hamiltonian
gives not only a back-reaction to the EHJ equation (\ref{EHJ eq.})
but also the geometric phase term
\begin{equation}
\left<\Phi_k(\phi,h_a) \right|
i \hbar \frac{\delta}{\delta \tau} \left|
\Phi_k ( \phi, h_a) \right>
= \frac{\hbar}{m_p^2} G_{ab} \frac{\delta S}{\delta
h_{(a}} \left( A_{b)} \right)_{kk}
\end{equation}
to the quantum state of the matter field.
It should also be noted that the gauge potential defined in terms of the
eigenstates of the generalized invariant does always give
nontrivial diagonal elements in strong
contrast with the gauge potential defined in terms of instantaneous
eigenstates of the Hamiltonian whose diagonal elements vanish for real
eigenstates.
On the other hand, when the gravitational
field Hamiltonian (\ref{gravitation Ham.})
acts on the wave function expanded by the eigenstates of the
generalized invariant as
\begin{equation}
\Psi (h_a, \phi) = \vec{U}^T (\phi, h_a) \cdot \vec{\Psi} (h_a),
\end{equation}
it becomes the
matrix nonadiabatic gravitational Hamiltonian equation and acquires
the induced gauge potential (\ref{gauge potential}) \cite{Kim1}
\begin{equation}
\hat{H}_g \vec{\Psi} (h_a) = \left[ \frac{1}{2m_p^2}
G_{ab} (\hat{\pi}^a -
A_a)(\hat{\pi}^b - A_b) -
2m_p^2c^2 \sqrt{h} ~^{(3)}R(h_a) \right]
\vec{\Psi}(h_a).
\label{nonad. grav. Ham. eq.}
\end{equation}
It was shown that there was a remarkable decoupling theorem
as mentioned earlier canceling the off diagonal gauge potential
and the expectation value of the matter field Hamiltonian
at the classical level. After the matter field Hamiltonian
is included, the total Hamiltonian $\hat{H}_g + \hat{H}_m$
acting on the $k$th wave function
\begin{equation}
\Psi_k (h_a,\phi) = \exp \left(\frac{i}{\hbar}S_k(h_a)\right)
 \left|\Phi_k (\phi,h_a) \right>
\label{kth wave func.}
\end{equation}
leads to
\begin{eqnarray}
\frac{1}{2m_p^2} G_{ab} \frac{\delta S_k}{\delta h_a}
\frac{\delta S_k}{\delta h_b}
&-& \frac{1}{m_p^2} G_{ab} \frac{\delta S_k}{\delta h_{(a}} ( A_{b)}
)_{kk} + \frac{1}{2m_p^2} G_{ab} (A_a)_{kk} (A_b)_{kk} \nonumber\\
 &-& 2m_p^2c^2 \sqrt {h} ~^{(3)}R(h_a) + \left<\Phi_k(\phi,h_a) \right|
 \hat{H}_m (\pi_\phi,\phi,h_a) \left|\Phi_k(\phi,h_a)\right> = 0.
\label{diag. nonad. eq.}
\end{eqnarray}
This equation can be rewritten as
\begin{eqnarray}
\frac{1}{2m_p^2}
G_{ab} \left(\frac{\delta S_k}{\delta h_a} - A_a \right)
\left(\frac{\delta S_k}{\delta
h_b} - A_b \right) - 2m_p^2c^2 \sqrt {h}
+ \left<\Phi_k(\phi,h_a) \right|
 \hat{H}_m (\pi_\phi,\phi,h_a) \left|\Phi_k(\phi,
h_a)\right> = 0.
\end{eqnarray}
Eq. (\ref{diag. nonad. eq.}) is the diagonal equation of Eq.
(\ref{matrix eq.}) with the wave function (\ref{kth wave func.}) and
the zeroth order solution $c_k = constant$.
In our new asymptotic expansion
of the WDW equation what corresponds to the gauge
potential is resigned to the matrix equation (\ref{matrix eq.})
for the matter field.

\section{MINIMAL FRW UNIVERSE}

We consider in detail the Friedmann-Robertson-Walker universe
minimally coupled to a free massive scalar field, whose
WDW equation is
\begin{equation}
\left[ \frac{\hbar^2}{2m_p^2} \frac{1}{a} \frac{\partial^2}{\partial a^2} - 2
m_p^2c^2 a - \hbar^2 \frac{1}{a^3} \frac{\partial^2}{\partial \phi^2} + m^2 a^3
\phi^2 \right] \Psi (a,\phi) = 0.
\label{FRW WD eq. 2}
\end{equation}
Here $m$ is the mass of the scalar field and the matter field Hamiltonian
is given by
\begin{equation}
\hat{H}_m = \frac{1}{a^3}\hat{\pi}_\phi ^2 + m^2a^3 \phi^2
\label{matter Ham.}
\end{equation}
In order to find the generalized invariant, the first thing to do
is to find the classical equation of motion for Eq. (\ref{matter Ham.})
\begin{equation}
\ddot{\phi}(\tau) + 3 \frac{\dot {a}(\tau)}{a(\tau)} \dot{\phi}
(\tau) + 4m^2 \phi (\tau) = 0,
\label{classical eq.}
\end{equation}
where the cosmological time will be determined later on
through Eq. (\ref{cosmological time}). The cosmological scale factor
$a$ and the scalar field $\phi$
depend implicitly on the cosmological time. Under the assumption
that the classical solutions, $\phi_1 (\tau)$ and
$\phi_2(\tau)$, to Eq. (\ref{classical eq.}) are given explicitly, it
is known that the generalized invariant is given by \cite{Cho}
\begin{equation}
\hat{I}(\tau) = g_-(\tau)\frac{\hat{\pi}_\phi^2}{2}
+ g_0(\tau) \frac{\hat{\pi}_\phi \hat{\phi} +\hat{\phi}
\hat{\pi}_\phi}{2}
+ g_+(\tau) \frac{\hat{\phi}^2}{2},
\label{generalized inv.}
\end{equation}
where
\begin{eqnarray}
g_-(\tau) &=& c_1 \phi_1^2 (\tau) + c_2 \phi_1 (\tau)\phi_2(\tau)
        +c_3 \phi_2^2(\tau),\nonumber\\
g_0(\tau) &=& -\frac{a^3(\tau)}{2}
\left( c_1 \phi_1 (\tau)\dot{\phi}_1(\tau)
 + \frac{c_2}{2}\left( \dot{\phi}_1 (\tau)\phi_2(\tau)+
  \phi_1 (\tau)\dot{\phi}_2(\tau)\right)
        +c_3 \phi_2(\tau)\dot{\phi}_2(\tau)\right),\nonumber\\
g_+(\tau) &=& \left(\frac{a^3(\tau)}{2} \right)^2
\left( c_1 \dot{\phi}_1^2 (\tau) + c_2 \dot{\phi}_1 (\tau)\dot{\phi}_2(\tau)
        +c_3 \dot{\phi}_2^2(\tau)\right).
\label{inv. rep.}
\end{eqnarray}

We may introduce the cosmological time-dependent
creation and annihilation operators of the generalized invariant
\begin{eqnarray}
\hat{b}^\dagger(\tau) &=& \left(\sqrt{\frac{\omega_0}{2g_-(\tau)}}-i\sqrt{
\frac{1}{2\omega_0 g_-(\tau)}}g_0(\tau) \right)\hat{\phi}-i\sqrt{\frac
{g_-(\tau)}{2\omega_0}}\hat{\pi}_\phi, \nonumber\\
\hat{b}(\tau) &=& \left(\sqrt{\frac{\omega_0}{2g_-(\tau)}}+i\sqrt{
\frac{1}{2\omega_0 g_-(\tau)}}g_0(\tau)\right)\hat{\phi}+i\sqrt{\frac
{g_-(\tau)}{2\omega_0}}\hat{\pi}_\phi,
\label{creation annihilation op.}
\end{eqnarray}
where
\begin{equation}
\omega_0 = \sqrt{g_+(\tau)g_-(\tau)-g_0^2(\tau)}
\end{equation}
is a constant of motion. The generalized invariant can be written as
\begin{equation}
\hat{I}(\tau) = \omega_0 \left(\hat{b}^\dagger(\tau)\hat{b}(\tau)
+ \frac{1}{2} \right).
\end{equation}
The eigenstates are the number states
\begin{equation}
\left|n,\tau \right> = \frac{1}{\sqrt{n!}} {\hat{b}}^{\dagger n}
(\tau) \left|0,\tau \right>,
\label{number state}
\end{equation}
where the ground state is annihilated by the cosmological time-dependent
annihilation operator
\begin{equation}
\hat{b}(\tau) \left|0,\tau \right> = 0.
\label{ground state}
\end{equation}
In the number state representation
the gauge potential (\ref{gauge potential}) reads that
\begin{equation}
A(\tau) = i\alpha(\tau) \left(\hat{b}^\dagger(\tau)\hat{b}(\tau) +
\frac{I}{2} \right) + \frac{i}{2} \left(\beta (\tau) \hat{b}^2(\tau)
-\beta^* (\tau){ \hat{b}}^{\dagger 2} \right),
\label{gauge potential rep.}
\end{equation}
where
\begin{eqnarray}
\alpha (\tau) &=& \frac{1}{2i} \frac{g_-(\tau)}{\omega_0}
\frac{\partial}{\partial \tau}
\left(\frac{g_0(\tau)}{g_-(\tau)} \right), \nonumber\\
\beta (\tau) &=& -\frac{1}{2} \frac{1}{g_-(\tau)}
\frac{\partial g_-(\tau)}{\partial \tau} + \alpha (\tau).
\end{eqnarray}
The gauge potential (\ref{gauge potential rep.}) has the matrix notation
\begin{eqnarray}
A_{kn}(\tau) &=& i \alpha (\tau) \left(n+ \frac{1}{2} \right)\delta_{k,n}
+ \frac{i}{2} \beta(\tau) \sqrt{n(n-1)} \delta_{k,n-2} \nonumber\\
&-& \frac{i}{2} \beta^* (\tau) \sqrt{(n+1)(n+2)} \delta_{k,n+2},
\end{eqnarray}
and the gauge potential squared
\begin{eqnarray}
A_{kn}^2 (\tau) &=&  \left(\left(-\alpha^2 +\frac{1}{2} \beta \beta^*
\right)(n^2 + n)
+ \frac{1}{2} \beta \beta^* -\frac{1}{4} \alpha^2
\right) \delta_{k,n} \nonumber\\
&-& \frac{1}{2} \alpha \beta (2n-1) \sqrt{n(n-1)} \delta_{k,n-2}
+ \frac{1}{2} \alpha \beta^* (2n+3) \sqrt{(n+1)(n+2)} \delta_{k,n+2}
\nonumber\\
&-& \frac{1}{4} \beta^2 \sqrt{n(n-1)(n-2)(n-3)} \delta_{k,n-4}
\nonumber\\
&+& \frac{1}{4} \beta^{* 2} \sqrt{(n+1)(n+2)(n+3)(n+4)} \delta_{k,n+4}.
\end{eqnarray}
One can show that
\begin{eqnarray}
\Omega_{kk}^{(1)}(\tau) &=& \hbar
 \frac{2\left(\omega_0^2+g_0^2(\tau) \right)+2m^2a^6g_-^2(\tau)}
 {2\omega_0a^3g_(\tau)} \left(k+\frac{1}{2}\right),
\label{FRW back-reaction}\\
\Omega_{kk}^{(2)}(\tau) &=& - \frac{\hbar}{2} \frac{g_-(\tau)}{\omega_0}
\frac{\partial}{\partial \tau}
\left(\frac{g_0(\tau)}{g_-(\tau)}\right) \left(k+\frac{1}{2}\right),
\label{FRW gauge potential}\\
\Omega_{kn}^{(3)} (\tau) &=& \frac{a}{\left( \frac{\partial S}{\partial a}
\right)^2} \left(i \frac{\partial A_{kn}}{\partial \tau} - A_{kn}^2
\right)
-i \frac{\partial}{\partial \tau} \left( \frac{a}{\frac{\partial S}
{\partial a}} \right) A_{kn}
\nonumber\\
&-& \frac{1}{a} \frac{\partial^2}{\partial a^2}
\delta_{kn}
- 2i A_{kn} \frac{\partial}{\partial a}.
\label{FRW gauge potential squared}
\end{eqnarray}
$\Omega $s are functions of the cosmological time
(\ref{cosmological time}) through
the dependence of the classical solutions (\ref{classical eq.}) and thereby
the generalized invariant (\ref{inv. rep.}) on $a(\tau)$. Therefore
Eq. (\ref{EHJ eq.}) which now reads
\begin{equation}
-\frac{1}{2m_p^2}\frac{1}{a} \left(\frac{\partial S}{\partial a} \right)^2
-2m_p^2 c^2a + \Omega_{kk}^{(1)}(\tau) = 0,
\label{FRW EHJ eq.}
\end{equation}
together with the definition of the cosmological time
\begin{equation}
\frac{\partial}{\partial \tau} = - \frac{1}{a} \frac{\partial S}
{\partial a} \frac{\partial}{\partial a},
\end{equation}
determines the action $S$ as a function of $a(\tau)$.

\section{Summary and Discussion}

In summary, we have developed a new asymptotic expansion method
according to which the Wheeler-DeWitt equation was separated
into the Einstein-Hamilton-Jacobi equation with the quantum back-reaction
of the matter field included and the time-dependent Schr\"{o}dinger
equation for the matter field. In the new asymptotic expansion
the Wheeler-DeWitt equation was equivalent
to two coupled nonlinear functional equations consisting
of the cosmological time
(\ref{cosmological time}) and of the
time-dependent Schr\"{o}dinger equation
(\ref{Schr eq.}) for the matter field or the matrix representation
(\ref{matrix eq.}) of it.
The time-dependent Schr\"{o}dinger equation
or the matrix equation for the matter field included quantum
gravitational corrections.
In particular we have found the exact quantum state (\ref{quantum state})
in the nonadiabatic basis of eigenstates
of the generalized invariant for the matter field Hamiltonian by
solving the matrix equation whose solution
consists of Eq. (\ref{lowest order sol.}) and
Eqs. (\ref{power series}) and ({\ref{series sol.}). The zeroth order
quantum state in Eq. (\ref{quantum state}) was the nonadiabatic basis
itself, which motivated the use of the generalized invariant to solve
the time-dependent Schr\"{o}dinger equation in Ref. \cite{Kim1}.
It was found that the quantum corrections of gravity
gave rise to transition of quantum states of the matter field
whose first order transition rate (\ref{first order quantum state})
was suppressed by the factor of
$1/m_p^2$, and the $n$th order transition rate by the factor of
$(1/m_p^2)^n$. Moreover, Eq. (\ref{lowest order sol.}) was
nothing but the Einstein-Hamilton-Jacobi
equation (\ref{EHJ eq.}) with the quantum back-reaction
of the matter field included.
The Einstein-Hamilton-Jacobi equation is an implicitly coupled
nonlinear equation for the gravitational action by which
the cosmological time (\ref{cosmological time})
used to define the generalized invariant is defined. Since
the quantum back-reaction of the matter field depends on
the mode-number, the cosmological time as well as
the gravitational action do depend the mode-number.
The physical meaning of the mode-dependent gravitational
actions is that the time-dependent Schr\"{o}dinger
equation (\ref{Schr eq.}) or its matrix equation
(\ref{matrix eq.}) should be defined along the
corresponding gravitational action. In fact
there are infinite number of wave functions which
are peaked along the gravitational action \cite{Kim2}.
The higher order quantum corrections of the gravity
to the matter field in Secs. III and V
are calculated along a specific mode-number gravitational action.

It has been shown through an investigation of the minimally coupled
Friedmann-Robertson-Walker universe
that the Einstein-Hamilton-Jacobi equation (\ref{EHJ eq.})
gives indeed the correct gravitational action rather than the conventional
Einstein-Hamilton-Jacobi equation (\ref{conv. EHJ eq.}) providing
an oscillatory regime necessary for the emergence of the Lorentzian universe.
The new asymptotic expansion method based on the generalized invariant
has the advantage that one has already canceled the off-diagonal
terms between two different eigenstates due to the remarkable decoupling
theorem of the generalized invariant, whereas one has to take care of the
off-diagonal terms between two different instantaneous eigenstates
of the matter field Hamiltonian in the conventional adiabatic expansion method.
Furthermore, by introducing the gauge potential (Berry connection) (\ref{gauge
potential}) we were able to express explicitly the gravitational correction
(\ref{second diff.}) as well as the back-reaction of the matter field.

Finally, we have applied the new asymptotic expansion method to
the Friedmann-Robertson-Walker universe with a minimal scalar field.
The generalized invariant was found in terms of the classical solutions
of the matter Hamiltonian, from which the back-reaction
(\ref{FRW back-reaction}), the geometric phase (\ref{FRW gauge potential}),
and the coupling matrix (\ref{FRW gauge potential squared}) of the
gravitational corrections are derived explicitly.

Considering an analogy, if any, between a quantum cosmological model of
gravitational field and matter fields
and a quantum system of heavy particles and light
particles, the gravitational field and heavy particles
behave as slow variables, and the matter fields and light particles
behave as fast variables \cite{Kim1}. In the quantum jargon of the fast
and slow variables the new asymptotic expansion method
provides us with a very systematic method to separate one equation
for the fast variables and the other equation for the slow variables.
The fast variables obey a parameter-dependent
quantum mechanical equation whose parameter is determined by the slow
variables and the slow variables satisfy a classical Einstein-Hamilton-Jacobi
equation with the quantum back-reaction of the fast variables.
In quantum cosmological models the gravitational field with the Planck
mass scale behaves as a heavy particle obeying classical
Einstein-Hamilton-Jacobi equation with the quantum back-reaction
of the matter fields which consists of the gauge potential as well as
the expectation value of the matter field Hamiltonian. The time-dependent
Schr\"{o}dinger equation for the matter field can also be solved
using the generalized invariant. The free massive scalar fields in
quantum cosmological models have been solved in Ref.\cite{Kim1} and
have provided a mechanism for the cosmological entropy production
during an expansion and recollapse of the Universe.
It is a new feature of  the new asymptotic expansion
of the Wheeler-DeWitt equation that
the geometric phases are an inevitable consequence to the quantum states
of the matter field as well as
the expectation value of the matter field Hamiltonian
to the Einstein-Hamilton-Jacobi equation
which is the counterpart of the classical
Einstein-Hamilton-Jacobi equation. The gauge potential in the nonadiabatic
basis can not be gauged away even for the real eigenstates.
However, the gauge potential
can always be gauged away for the adiabatic basis of real eigenstates of the
matter field Hamiltonian
in the conventional expansion method. Thus the argument \cite{Kim1}
that the cosmological time asymmetry may have origin in the geometric phase
still survives.

\acknowledgments
The author would like to thank the Russian Gravitational Association for
warm hospitality during the International School-Seminar on the
Multidimensional
Gravity and Cosmology where part of this work was done and would like to thank
the Korea Science and Engineering Foundation, 1994 for supporting the
participation.
This work was supported by the Non-Directed Research Fund, Korea Research
Foundation, 1994.


\begin{references}
\bibitem[*]{Kim} Electronic address: sangkim@knusun1.kunsan.ac.kr

\bibitem{Banks} T. Banks, Nucl. Phys. {\bf B249}, 332 (1985).
\bibitem{Gerlach} U. H. Gerlach, Phys. Rev. {\bf 177}, 1929 (1969).
\bibitem{Brout1} R. Brout, Found. Phys. Phys. {\bf 17}, 603 (1987);
R. Brout and D. Weil, Phys. Lett. B {\bf 192}, 318 (1987); R. Brout,
Z. Phys. B {\bf 68}, 339 (1987).
\bibitem{Kiefer1} C. Kiefer, Class. Quantum Grav. {\bf 4}, 1369 (1987);
C. Kiefer and T. P. Singh, Phys. Rev. D {\bf 44}, 1067 (1991).
\bibitem{Singh} T. P. Singh and T. Padmanabhan,
Ann. Phys. {\bf 196} 296 (1989); T. P. Singh,
Class. Quantum Grav. {\bf 7}, L149 (1990)
\bibitem{Paz} J. P. Paz and S. Sinha,
Phys. Rev. D {\bf 44}, 1038 (1991).
\bibitem{Gundlach} C. Gundlach, Phys. Rev. D {\bf 48}, 1700 (1993).
\bibitem{Kiefer2} C. Kiefer, R. M\"{u}ller, and T. P. Singh,
{\it Quantum gravity and non-unitarity in black hole evaporation},
gr-qc/9308024 (1993).
\bibitem{Balbinot} R. Balbinot, A. Barletta, and G. Venturi, Phys. Rev. D
{\bf 41}, 1848 (1990).
\bibitem{Brout2} R. Brout and G. Venturi,
Phys. Rev. D {\bf 39}, 2436 (1989).
\bibitem{Datta} D. P. Datta, Mod. Phys. Lett. {\bf A8}, 191 (1993); {\it
ibid} {\bf 8}, 2523(E) (1993); {\it ibid} {\bf 8}, 601 (1993); Phys. Rev.
D {\bf 48}, 5746 (1993); {\it ibid} {\bf 49}, 2605 (1993); {\it Relevance
of induced gauge interactions in decoherence}, IC/94/168 (1994) (to appear
in Phys. Rev. D). {\it Can vacuum energy gravitate?}, (1994) (to appear in
Gen. Rel. Grav.)
\bibitem{Kim1} S. P. Kim, {\it Time asymmetry originated from
geometric phases in quantum cosmology} in Proceedings of The Second
Haengdang Summer Workshop on Theoretical Physics edited by H. K. Lee
(Hanyang Univ., Seoul, 1993); S. P. Kim and S.-W. Kim,
Phys. Rev. D {\bf 49}, R1679 (1994); S. P. Kim, Phys. Lett. A {\bf 191},
365 (1994); S. P. Kim and S.-W. Kim,
Phys. Rev. D {\bf 51}, (1995) (in press).
\bibitem{Kiefer3} For a good review of semiclassical gravity both with
the back-reaction and with the gauge potential see
C. Kiefer, {\it The semiclassical
approximation to quantum gravity} in {\it Canonical gravity
 - from classical to quantum}, edited by
J. Ehlers and H. Friedrich (Springer, Berlin, 1994),
and references are therein.
\bibitem{Kim2} S. P. Kim, Phys. Rev. D {\bf 46}, 3403 (1992);
S. P. Kim, J. Kim , and K. S. Soh, Nucl. Phys. {\bf B406}, 481 (1993).
\bibitem{Lewis} H. R. Lewis, Jr. and W. B. Riesenfeld, J. Math. Phys.
{\bf 10}, 1458 (1969).
\bibitem{Halliwell}J. J. Halliwell and S. W. Hawking, Phys. Rev. D
{\bf 31}, 1777 (1985).
\bibitem{Kiefer4} C. Kiefer, Phys. Rev. D {\bf 38}, 1761 (1988).
\bibitem{Cho} K. H. Cho and S. P. Kim, J. Phys. A: Math. Gen. {\bf 27},
1387 (1994); S. P. Kim, {\it ibid} {\bf 27}, 3927 (1994);
J. Y. Ji, J. K. Kim, and S. P. Kim, Phys. Rev. A {\bf 51}, 4268 (1995).



\end{references}
\end{document}